\documentclass{aastex}
\usepackage{emulateapj5}
\usepackage{epsfig}

\makeatletter
\newenvironment{inlinefigure}{%
\def\@captype{figure}%
\noindent\begin{minipage}{0.999\linewidth}\begin{center}}
{\end{center}\end{minipage}\smallskip}
\makeatother
\newcommand{\msun}{M_\odot}
\newcommand{\msunyr}{\msun {\rm yr}^{-1}}
\newcommand{\rl}{R_L}
\newcommand{\dd}{{\rm d}}

\submitted{Accepted for publication in ApJ (August 2001)}

\shorttitle{Kolb et al.}
\shortauthors{Nova--induced mass transfer variations}

\begin{document}

\title{Nova--induced mass transfer variations}


\author{U. Kolb}
\affil{Department of Physics and Astronomy, The Open University, \\
Walton Hall, Milton Keynes MK7 6AA, UK}
\email{U.C.Kolb@open.ac.uk}

\bigskip

\author{S. Rappaport}
\affil{Department of Physics and Center for Space Research \\
Massachusetts Institute of Technology, Cambridge, MA 02139}
\email{sar@mit.edu}

\bigskip

\author{K. Schenker}
\affil{Department of Physics \& Astronomy, University of Leicester, \\
University Road, Leicester LE1 7RH, UK}
\email{kjs@star.le.ac.uk}
\and
\author{S. Howell
}
\affil{Astrophysics Group, Planetary Science Institute \\
620 N.~6th Avenue, Tucson, AZ  85705}
\email{howell@psi.edu}


\begin{abstract}
We investigate variations of the mass transfer rate in cataclysmic
variables (CVs) that are induced by nova outbursts.
The ejection of nova shells leads to a spread of transfer rates
in systems with similar orbital period.
The effect is maximal if the specific
angular momentum in the shell is the same as the specific
orbital angular momentum of the white dwarf. We show analytically that
in this case the nova--induced widening of the mass transfer rate distribution
can be significant if the system, in the absence of nova outbursts, is close
to mass transfer instability (i.e., within a factor of $\sim 1.5$ of the
critical mass ratio).
Hence the effect is negligible below the period gap and for systems with
high--mass white dwarfs. At orbital periods between about 3 and 6~hrs
the width of the mass transfer rate distribution exceeds an order of magnitude
if the mass accreted on the white dwarf prior to the runaway is larger
than a few $10^{-4} \msun$.
At a given orbital period in this range, systems with the highest
transfer rate should on average have the largest ratio of donor to
white dwarf mass. We show results of population
synthesis models which confirm and augment the analytic results.
\end{abstract}

\keywords{novae, cataclysmic variables
$-$ stars: binaries: close
$-$ stars: evolution
$-$ stars: mass loss
$-$ stars: low mass}

\section{Introduction}

A long--standing problem in the theory of cataclysmic variable (CV)
evolution is the large scatter of observationally deduced mass transfer
rates for systems with similar orbital periods (Patterson 1984, Warner 1987,
Warner 1995). This appears to be in conflict with
the well founded assumption that mass transfer in CVs is driven by
orbital angular momentum losses such as gravitational radiation and
magnetic braking (see, e.g., Rappaport, Joss, \& Webbink 1982 (RJW),
Rappaport, Verbunt, \& Joss 1983 (RVJ), Hameury et al. 1988;
for a review see, e.g.,\ King 1988).

Binary evolution calculations have shown that CVs that form with
different initial secondary masses converge rather quickly to an essentially
unique track in, e.g.,\ the orbital period -- mass transfer rate
($P_{\rm orb}-\dot M$) space (e.g., RVJ, \ Kolb \& Ritter 1992).
This track convergence arises as the angular momentum loss rate,
and hence the mass transfer rate itself, depend on the
current system parameters only. Low--mass main--sequence stars that
are subject to mass loss rapidly settle at a new equilibrium radius $R_2$
(Stehle et al.\ 1996) which in turn is a function of the mass loss
timescale. Therefore the orbital period $P_{\rm orb} \propto R_2^{3/2}$
of systems at the equilibrium radius is set by the mass transfer rate.

The track convergence implies that any variation of $\dot M$ at a
fixed $P_{\rm orb}$ should largely reflect the dependence of the angular
momentum loss rate $\dot J$ on the white dwarf mass $M_1$. Since
$-\dot M_2/M_2 \propto \dot J/J$, cf.\ eq~(\ref{Xstat}) below, and
$\dot J$ has no explicit dependence on $M_1$ for mechanisms like
magnetic braking which are rooted in the secondary star, the expected
dependence on $M_1$ at a fixed period is $\propto
(M_1+M_2)^{1/3}/M_1$. This is too weak in view of the observed
scatter of at least an order of magnitude.

Classical nova outbursts represent an obvious mechanism that could cause
deviations of the transfer rate from the secular mean value set
by the angular momentum loss rate. All CVs are expected to experience
some form of nuclear burning of the accreted hydrogen--rich material
on the surface of the white dwarf (WD). In most cases this should be
non--steady and lead to systemic mass loss, e.g. in the form of a
classical nova outburst.  Specifically, in the following we assume
that the burning is episodic and always accompanied
by the ejection of the layers that are affected by the burning, i.e.,
the nova shell.  The ejection occurs on a very short
timescale ($\la$ months) and effectively leads to an instantaneous
change of the orbital separation. The magnitude and sign of the change
are determined by the amount of angular momentum the shell removes from
the orbit. In most cases the orbital separation increases, with a resulting
drop of the mass transfer rate, as the Roche lobe is moved suddenly
outward by a few atmospheric pressure scale heights from the photosphere.
The effect is largest if the angular momentum per unit mass in the nova
shell is minimal. For an axi-symmetric ejection this minimum is just the
specific orbital angular momentum of the
white dwarf. We refer to such hydrodynamic ejections as {\em ballistic},
irrespective of the physical nature of the ejection mechanism (e.g., nova
explosions, optically thick winds from the white dwarf, and so forth).

If there is additional angular momentum transfer from the binary orbit to the
nova envelope due to dynamical friction of the secondary orbiting inside this
envelope the increase of the orbital separation upon envelope ejection is
smaller than in the ballistic case. Very strong frictional transfer would
even lead to a decrease of the separation.

The approach taken by MacDonald (1986) to model dynamical friction in the
nova envelope gave an extremely efficient extraction of
angular momentum from the orbit. The concomitant mean orbital evolution
was greatly accelerated and resulted in both very large mean mass transfer
rates
($\ga 10^{-7} \msunyr$), and a greater spread of the mean mass
transfer rate at a given orbital period than in the absence of nova outbursts.
Such a variation of the {\em mean} transfer destroys the
coherence of evolutionary tracks we noted above, and is therefore
in conflict with the standard model for the CV orbital period gap
(see e.g. Kolb 1996). MacDonald did not consider the effect of individual
outbursts on the mass transfer rate distribution.

In contrast, Schenker, Kolb, \& Ritter (1998), following Livio,
Govarie, \& Ritter (1991) to model dynamical friction, found a much weaker
angular momentum transfer by dynamical friction. Schenker et al.\ largely
dismissed the effect of nova outbursts on the mass transfer rate
distribution as
unimportant, for ballistic ejections as well as for cases with plausible
friction.  Note that for the remainder of this paper we use the terms ``mass
transfer rate distribution'' and ``mass transfer rate spectrum''
interchangeably.
More precisely, they both describe the distribution in values of $\dot M$
at a given orbital period, for either the evolution of a single CV or an
ensemble of CVs, as a function of orbital period.

A closely related issue is the notion that CVs ``hibernate'' after
nova outbursts, i.e.,\ there has been the suggestion that post--novae are
very faint, for a time comparable to the nova recurrence time (Shara
et al.\ 1986; Prialnik \& Shara 1987, Livio \& Shara 1987). The
hibernation idea was motivated by space density issues, the need for
smaller mass transfer rates to obtain violent nuclear runaways in
theoretical nucleosynthesis models, and by
observations suggesting that the remnant systems of historical novae are faint.

In this work we reconsider the maximum possible variation of the mass transfer
rate about the secular mean rate due to the ballistic ejection of nova shells.
In Section 2 we review the methods by which
the mass transfer rates in CVs are deduced.  We conclude that while some
of the spread in the $\dot M$'s is due to the uncertainties in determining the
transfer rates, much of the spread is due to a real variation among
different systems.  In Section 3 we develop a simple analytical description
for the time--dependent behavior of $\dot M$ between nova outbursts, as well
for the probability densities for $\dot M$ at a given orbital period (with
technical details deferred to the Appendix).   In this same section
we apply the analytic model to find the previously ignored optimum
system parameter combination that maximizes the spread of mass
transfer rates at a given orbital period in an observed sample of CVs.
In Section 4 we use this prescription in population synthesis models
to show the overall effect on a theoretical CV population with
standard assumptions about their formation and evolution.
A significant fraction of the CV population lies in the optimum
parameter range found in Section 3.
Finally in Section 5 we draw conclusions
concerning the likely effects of nova explosions on the spread in
values of $\dot M$ observed in CVs.

\section{Determination of $\dot M$ in CVs}

Observational determinations of the mass accretion rate in cataclysmic
variables show a large spread of order 10 up to 100 in $\dot M$ at a given
orbital period, being most pronounced in the period range of 3 to 6 hr
(see Fig. 9.8 in Warner 1995).  Before presenting our model which may
partially account for this substantial spread we briefly review the
observational methods used to determine $\dot M$ and examine what ``spread"
may be introduced simply by uncertainties in the measurements or the
assumptions that go into the analysis and interpretation of the data.

Smak (1989) computed $\dot M-M_V$ relationships for standard thin
accretion disks using the usual multi-temperature black-body approximations.
These relations are still often used for determinations of $\dot M$
(see, e.g.,\ Robinson et al.\ 1999). Smak (1994) later added
a finite disk thickness to account for a bright disk rim and a less oblique
view of the far side of the disk.  He found that M$_V$ increased by only 0.5
magnitudes in most cases for moderate to low binary inclinations $i$, while
at large values of $i$ and/or high transfer rates, changes in M$_V$ of
up to 2.5 magnitudes are possible.  Smak's results show that for high
mass accretion rates ($10^{-8}-10^{-9}~M_{\odot}$ yr$^{-1}$)
a perfectly known value of M$_V$  corresponds to an uncertainty in
accretion rate of a factor of $\sim 2-10$ at a given orbital period.

To make use of the $\dot M-M_V$ relations, the value of M$_V$ for the
disk is needed. Observational determinations of M$_V$ are usually
based upon the belief that in the optical V band, the flux received is
almost entirely from the accretion disk.  For CVs with orbital periods
in the range of $3-6$ hr, this is a fairly safe assumption.
Thus, using some method of distance determination,
and accounting for the binary inclination (see Warner 1995 for the convention
used to convert the observed value into a common definition of M$_V$), one
can then estimate $\dot M$.  To the uncertainties inherent in the $\dot M -
M_V$ relationships one should also add the rather substantial uncertainties in
the distance determinations which are used to find M$_V$.

An independent method of checking whether these $\dot M$ estimates
are reasonable, which is not based on the determination of
M$_V$, is to utilize X-ray observations and simple accretion disk/boundary
layer theory.  Here the basic assumption is that some fraction of the
accretion luminosity, $GM_1\dot M/R_1$, will emerge in
X-radiation.  Thus,
if the mass $M_1$ and radius $R_1$ of the white dwarf are known to some
degree from observation or by assumption,
measurements of $L_X$ can be used to estimate $\dot M$ (see Frank et
al. 1992). Patterson and Raymond (1985a) provided a comparison of
observed $L_X$ with a suitable observed mean of $\dot M$ as a function
of orbital period, derived from $\dot M$ estimates by other methods
(e.g.\ Patterson 1984), and found the following:
Using an optically thick boundary layer model the
two estimates of $\dot M$ agreed to within a factor of 5 at low $\dot M$
($10^{-10.5}~M_{\odot}$ yr$^{-1}$) and 10 at high $\dot M$
($10^{-8.5}~M_{\odot}$ yr$^{-1}$).

Another powerful method to estimate $\dot M$ is photometric eclipse
mapping. This uses the standard relation $T^4(R) \simeq 3 G M_1\dot
M/8\pi r^3$ between disk temperature $T$
and disk radius $r$ to allow an estimate of $\dot M$ (Horne 1985). If
multicolor light curves are available, and the effect of reddening can
be estimated, this method is independent of distance determinations.
Smak (1994) showed that for $\dot M$ $\ga$ $10^{-8} M_{\odot}$
yr$^{-1}$ (orbital periods of greater than $\sim 5$ hr), this
technique can yield spurious results.
However, for typical systems in the $3-6$ hr range,  the agreement between
the observed and theoretically expected $T^4(R)$ profiles
is generally good, with deviations being a factor of 10 or less.
Baptista (2000) reviews the current level of knowledge in this area.

Yet another method, using equivalent widths of disk emission lines,
is based on apparent relationships between the strength of the disk
emission lines and M$_V$ from the accretion disk.
For example, Patterson (1984) used the equivalent width of H$\beta$
to present a set of accretion disk $\dot M$ values.
Smak (1989), however, noted that Patterson used a mixture
of thin and thick accretion disks at various binary inclinations and
his assumed $\dot M$ values were based on an $\dot M-M_V$ relation by
Tylenda (1981) which, in turn, used an unrealistically large
accretion disk. Thus, the determined $\dot M$ values are likely to be
underestimated in general by factors of $2-6$.
Using the strength of
He II emission, an assumed white dwarf mass of 0.7 $M_{\odot}$,
and a 3 \AA ~equivalent width for the line, Patterson and Raymond (1985b)
found agreement between the values of $\dot M$ predicted by their
method, and the values of $\dot M$ estimated by other methods
to within factors of $5-10$ over a broad range of $\dot M$ for a
number of CVs.

We find that, in general, the observational determinations of
$\dot M$ have typical associated uncertainties of factors of $\sim 3-5$ and in
some cases up to a factor of $\sim 10$. From this information we would
conclude that the observed spread in the $\dot M$ distribution, which is
found to approach factors of 100, especially for the $3-6$ hr orbital
period range, is in large part a real effect.

In the final analysis, however, perhaps the best evidence in favor of a wide
spread in the values of $\dot M$ at a given $P_{\rm orb}$, is the coexistence
of dwarf novae and nova-like variables at the same orbital period.  The widely
accepted explanation for dwarf nova eruptions is provided by thermal ionization
instabilities in their accretion disks
(Meyer \& Meyer--Hofmeister 1981; Cannizzo, Shafter, \& Wheeler 1988;
for a review see e.g.\ Frank et al.\ 1992)
which, in turn, require smaller values of $\dot M$ than are
inferred for nova-like variables which exhibit no such disk instabilities.

\section{Analytic model}

In this section we develop an analytic model for the mass transfer
rate spectrum that is induced by the ballistic ejection of nova shells.
For convenience we defer a portion of the derivation to the
Appendix. After introducing standard definitions describing mass
transfer via Roche--lobe overflow we simply present
the main resulting expressions and focus on their discussion.

\subsection{Stationary and nonstationary mass transfer}

For simplicity of notation, let $\dot M$ be the magnitude of the mass transfer
rate, i.e.,\
\begin{equation}
\dot M = - \dot M_2 > 0,
\end{equation}
where $M_2$ is the mass of the Roche--lobe filling secondary. In
general, the transfer rate is a sensitive function of the difference
between the stellar (photospheric) radius $R$ and the Roche lobe
radius $\rl$ of the secondary. In the case of CVs a good approximation
is
\begin{equation}
\label{mdot}
\dot M= \dot M_{ph} \, \exp \left\{\frac{R-\rl}{H}\right\}.
\end{equation}
Here $H$ is the photospheric pressure scale height of the
secondary star, and $\dot M_{ph}$ is the mass transfer rate when the Roche
lobe is situated at the photospheric radius
%
%
(see, e.g.,\ Ritter 1988, Kolb \& Ritter 1990).   Typical values of
$\dot M_{ph}$ for
low-mass main-sequence stars are expected to be $\simeq 10^{-8} \msunyr$.

Over intervals that are short compared to the mass loss time scale
$t_M=M_2/\dot M$, the time derivative of the transfer rate (\ref{mdot}) is
\begin{equation}
\frac{d\dot M}{dt} \simeq \dot M \, \frac{R}{H} \, \left\{ \frac{\dot R}{R} -
\frac{\dot \rl}{\rl} \right\}
\label{Xdot0}
\end{equation}
if $\dot M_{ph}$ and $H$ are constant and $R-\rl \ll R$.

To make use of (\ref{Xdot0}) we introduce the standard decomposition
of radius derivatives into mass transfer--dependent and independent
terms (e.g.\ RVJ, Webbink 1985, Ritter 1996),
\begin{equation}
\frac{\dot R}{R} = \zeta \, \frac{\dot M_2}{M_2} + \left(\frac{\partial
\ln R}{\partial t} \right)_{\rm th}
\label{Rdot}
\end{equation}
\begin{equation}
\frac{\dot \rl}{\rl} = \zeta_L \, \frac{\dot M_2}{M_2} +  \, 2 \left(
\frac{\dot
J}{J} \right)_{\rm sys}.
\label{Rldot}
\end{equation}
Here the stellar mass--radius index $\zeta$ describes the adiabatic
response of the donor star to mass loss, while $(\partial \ln
R/\partial t)_{\rm th}$ denotes the
relative radius change due to thermal relaxation. (We neglect any
changes of the radius due to nuclear evolution).
$\zeta_L$ is the Roche--lobe index and encompasses changes of the Roche
lobe due to mass transfer and mass loss, and
$\dot J_{\rm sys}$ are ``systemic'' orbital angular momentum losses
unrelated to mass loss (e.g., due to gravitational radiation and magnetic
braking). $J$ is the total orbital angular momentum.

With (\ref{Rdot}) and (\ref{Rldot}) equation~(\ref{Xdot0})
can be written as
\begin{equation}
\frac{d\dot M}{dt} =  \frac{1}{\epsilon} \, \frac{\dot M}{M_2} \, \left\{
\frac{M_2}{t_{\rm
ev}} - D \dot M \right\} ,
\label{Xdot1}
\end{equation}
where we have introduced the relative thickness of the photosphere of the
donor star,
\begin{equation}
\epsilon = H/R,
\end{equation}
(for CVs, $\epsilon \simeq 10^{-4}$), the stability denominator
\begin{equation}
D = \zeta - \zeta_L,
\label{D}
\end{equation}
usually of order unity, and the evolutionary timescale $t_{\rm ev}$,
\begin{equation}
\frac{1}{t_{\rm ev}} = \left(\frac{\partial
\ln R}{\partial t} \right)_{\rm th} \, - \, 2 \left( \frac{\dot J}{J}
\right)_{\rm sys}.
\end{equation}

From (\ref{Xdot0}) we see that mass transfer is stationary (d$\dot M$/dt
=0) if
\begin{equation}
\frac{\dot R}{R} = \frac{\dot \rl}{\rl},
\end{equation}
and (\ref{Xdot1}) shows that in stationarity the mass transfer rate is
\begin{equation}
\dot M_s = \frac{M_2/t_{\rm ev}}{D}.
\label{Xstat}
\end{equation}
Hence the transfer rate is of order $M_2/t_{\rm ev}$, unless the
system is close to instability where $D$ is small (see e.g.\ RJW, RVJ, King \&
Kolb 1995 for a discussion of mass transfer instability).

\subsection{Nova outbursts}

To apply this formalism to CVs that experience repeated nova
outbursts we consider both the actual evolution between outbursts,
and a mean evolution.


We assume that the evolution between outbursts is conservative, i.e., none
of the transferred mass is lost from the system.

The outburst itself is treated as an instantaneous ejection of the
nova shell with mass $\Delta M_{\rm ej}$.
If this is related to the accreted mass $\Delta M_{\rm ign}$ needed to
trigger the thermonuclear runaway as
\begin{equation}
\Delta M_{\rm ej} \, = \, \gamma \Delta M_{\rm ign} \, = \, - \gamma
\Delta M_2 > 0,
\label{eq:gamma}
\end{equation}
($\gamma>0$), then the ratio of maximum to minimum values of $\dot M$ during an
outburst cycle is
\begin{equation}
\frac{\dot M_f}{\dot M_0} = \exp \left( \frac{1}{\epsilon} \, \gamma \,
\frac{4}{3} \,
\frac{\Delta M_{\rm ign}}{M_b} \right) =
\exp \left( \frac{1}{\epsilon} \, \frac{4}{3} \, \frac{\Delta M_{\rm
ej}}{M_b} \right),
\label{A3}
\end{equation}
where $\dot M_f$ and $\dot M_0$ are the values of $\dot M$ just before and just
after a nova explosion, respectively, and $M_b$ is the total mass of the
binary. Equation~(\ref{A3}) assumes that the ejected mass carries away the mean
specific orbital angular momentum of the white dwarf. In this case
the Roche lobe of the secondary always increases as a result of a nova
outburst, leading to a sudden drop of the mass transfer rate.

We assume that the nova outburst recurrence time $t_{\rm rec}$ is
sufficiently short so that the donor mass $M_2$, the relative
thickness $\epsilon =H/R$ of the photosphere and the evolutionary
timescale $t_{\rm ev}$ can be considered as constant between outbursts.
Then, by direct integration of eq. (6), the transfer rate
between $t=0$, the time immediately after the outburst, and $t=t_{\rm
rec}$, the time immediately before the next outburst, is
\begin{equation}
\dot M(t) = \frac{\dot M_c}{1 + (\dot M_c/\dot M_0 -1)\exp(-t/\epsilon
t_{\rm ev})}.
\label{Xt}
\end{equation}
(see, e.g.,\ Schenker et al.\ 1998). Here $\dot M_c$ is the stationary mass transfer rate the system would adopt (for conservative transfer) in the
absence of nova outbursts, except when it is dynamically unstable. A
strict definition of $\dot M_c$ is given in the Appendix, see
equation~(\ref{Xc}).

\subsection{Mass transfer rate spectrum}

Consider now an ensemble of CVs with similar component masses ($M_1$,
$M_2$) and mean mass transfer rate $\dot M_m$ at orbital period
$P$. The mean transfer rate is defined by
\begin{equation}
\dot M_m = \frac{1}{t_{\rm rec}} \int_0^{t_{\rm rec}} \dot M(t) \dd t
\label{trec}
\end{equation}
and corresponds to the ``fiducial" mean evolution where the sequence of
discontinuous mass ejections is replaced by a continuous, isotropic
wind, such that $\dot M_{\rm wind} = \Delta M_{\rm ej}/t_{\rm rec}$.

Then the actual (instantaneous) mass transfer rates in the
ensemble are between $\dot M_0$ and $\dot M_f = \dot M(t_{\rm rec})$.
The probability density distribution $n(\dot M)$ in $\dot M$ space is
proportional to $1/\dot M$, so that the normalized mass transfer rate
spectrum reads
\begin{equation}
n(\dot M) = \frac{1}{t_{\rm rec}} \, \frac{\dd t}{\dd \dot M}(\dot M)
\label{n}
\end{equation}
($\int n(\dot M)\dd \dot M=1$).
With $\dd t/\dd \dot M = 1/(d\dot M/dt)$ we obtain by manipulating
equation~(\ref{Xdot1}):
\begin{equation}
\frac{\dd t}{\dd \dot M} = \epsilon \, t_{\rm ev} \, \frac{\dot M_c}{\left(
                      \dot M_c-\dot M \right) \, \dot M}.
\label{dtdX}
\end{equation}

As can be seen from (\ref{A3}) the full ratio of mass transfer rates
exceeds an order of magnitude if the ejected mass is larger than about
$10^{-4} \msun$. Yet, in many cases, the systems still spend most of
their time close to the secular mean $\dot M_m$, even if the amplitude
is large. The corresponding mass transfer rate spectrum
$\tilde{n}(\log \dot M) \propto \dot M
n(\dot M)$ has a narrow spike at large transfer rates, and a long tail with low
probability density toward small $\dot M$.  An illustrative example
of $\dot M$ vs. time during the interval between nova events is shown
in the upper panel of Fig. 1.  The lower panel in the same figure shows
the corresponding probability density in $\dot M$.
The cases studied by Schenker et al.\ (1998) had an
even narrower distribution than the example shown in Fig.~1.

To judge the effective ``width'' of the distribution in $\dot M-$space
we consider the half width $w$ to be the following ratio:
\begin{equation}
w \equiv \frac{\dot M_f}{\dot M_{1/2}},
\label{ww}
\end{equation}
where $\dot M_{1/2} = \dot M(t_{\rm rec}/2)$ is the median of the mass transfer
rate distribution (\ref{n}), i.e.\
\begin{equation}
\int_{\dot M_0}^{\dot M_{1/2}} n(\dot M) \dd \dot M = \int_{\dot
M_{1/2}}^{\dot M_f} n(\dot M)
\dd \dot M.
\end{equation}
As a working criterion, we adopt the condition $w>10$ to signal a
``wide distribution''.  If we define a ``stability" ratio
\begin{equation}
\alpha = \frac{2 D_c}{D_m}
\end{equation}
and a mass accumulation factor
\begin{equation}
x = \exp \left( \frac{D_m}{2\epsilon} \, \frac{\Delta M_{\rm
ign}}{M_2} \right)
\end{equation}
($x\geq1$), then equations (\ref{Xt}) and (\ref{ww}) and the above
definitions can be combined to yield the following expression for $w$:
\begin{equation}
w(x) = \frac{x (1+x^{1-\alpha})}{1+x}.
\label{wx}
\end{equation}
The stability denominators $D_c$ and $D_m$ are given by (\ref{D}) and
refer to the conservative evolution between outbursts and the
time--averaged evolution, respectively. The stellar index is the same
for both $D_c$ and $D_m$, while the appropriate Roche lobe indices
are given in the Appendix.

\begin{inlinefigure}\label{fig1}
\plotone{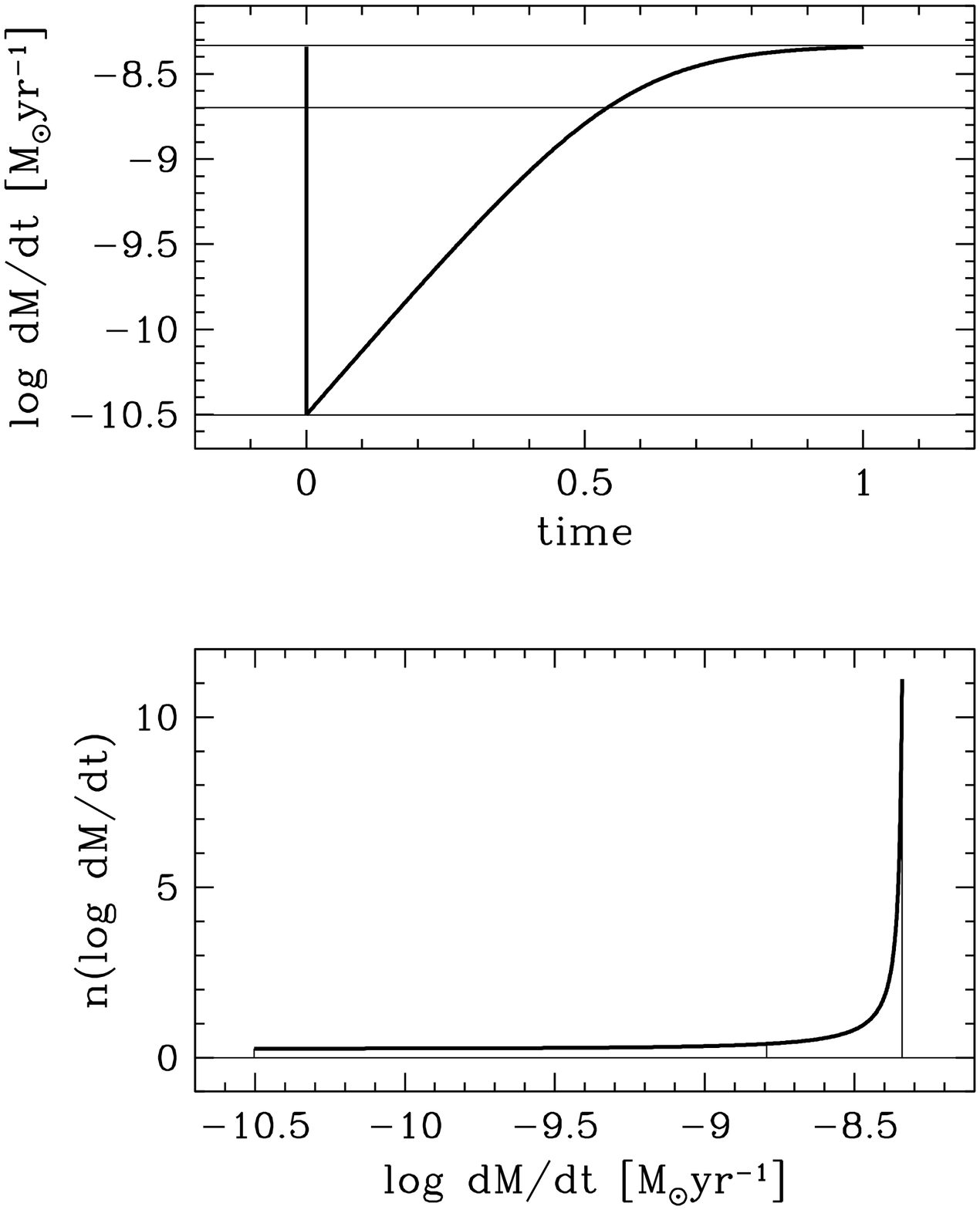}
\caption{{\em Upper panel:} Typical evolution of the mass
transfer rate $\dot M$ between nova outbursts (here for $M_2=0.3\msun$,
$M_1=0.6\msun$, $\Delta M_{\rm ign}=3 \times 10^{-5} \msun$).
The time is in units of the recurrence time. The upper solid line is
the asymptotic value $\dot M_c$, the lower solid line the time--average
$\dot M_m=2\times10^{-9}\msunyr$.
{\em Lower panel:} The
corresponding probability density distribution $n(\log \dot M)$.}
\end{inlinefigure}

As $\zeta$ is essentially a function of stellar mass, and $\zeta_L$ is
a function of the mass ratio $q$,
\begin{equation}
q= \frac{M_2}{M_1},
\end{equation}
only, the stability factor $\alpha$ is
fixed for a given system.
In contrast, the mass
accumulation factor $x$ depends on the ignition mass.
Observational estimates of nova ejecta masses,
though subject to large uncertainties,
tend to give much larger values
than results of detailed model calculations of the thermonuclear runaway
on the surface of hydrogen--accreting white dwarfs
(see, e.g.,\ Starrfield 1999 and references therein). Hence we examine
$w$ as a function of $x$ in order to provide possibly new insight into
nova ejecta masses based on the spread in CV mass transfer rates.

\begin{inlinefigure}\label{fig3}
\plotone{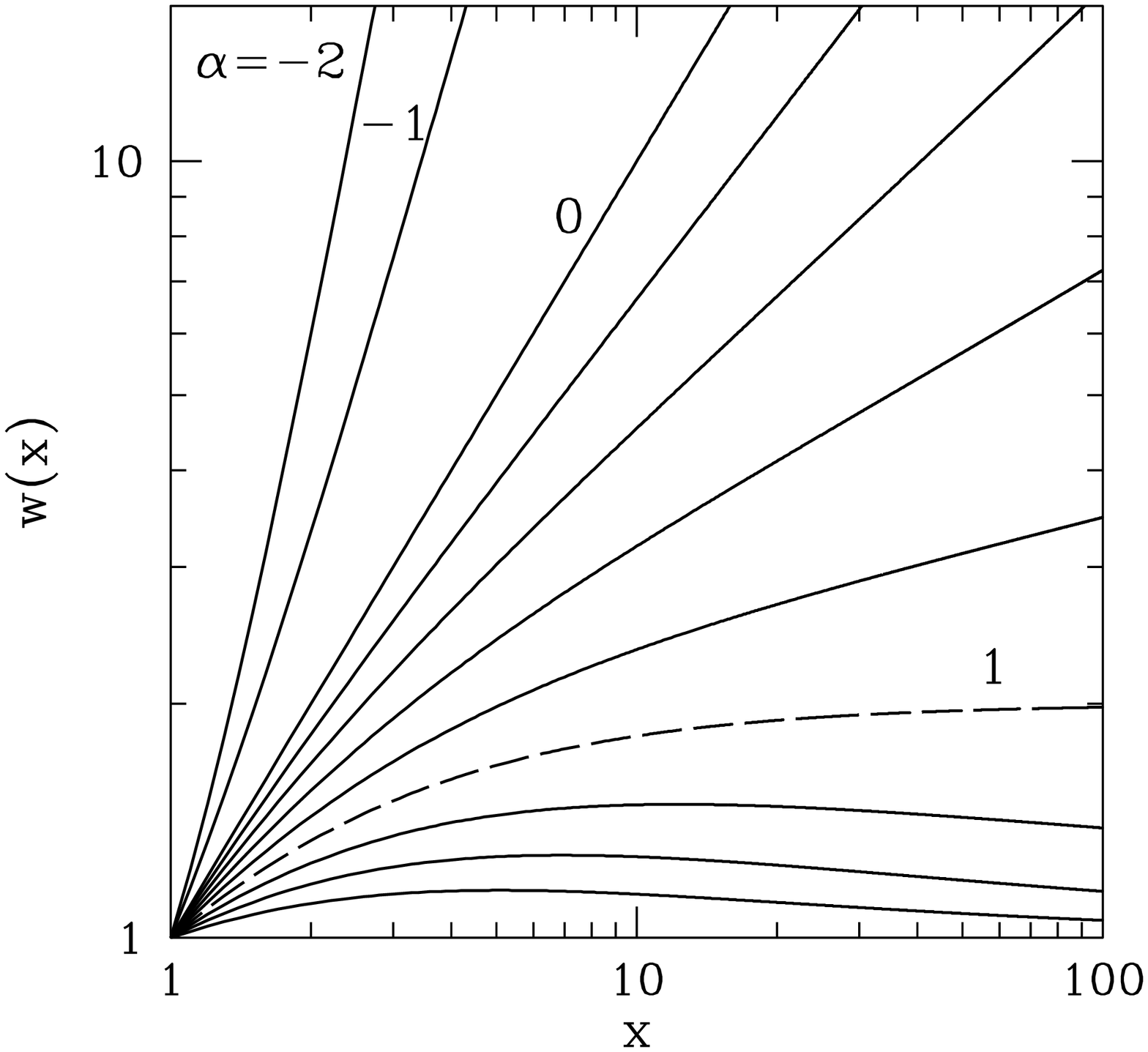}
\caption{Half width $w(x)$ as a function of mass accumulation
factor $x$, for different stability ratios $\alpha$ (as labelled). The
dashed curve is the limiting case $\alpha=1$. The $\alpha$ increment
between the curves is 0.2 if $\alpha>0$.}
\end{inlinefigure}

The shape of $w(x)$ depends on the value of the parameter $\alpha$
which describes the stability of mass transfer in the system.
We require that CVs are dynamically stable with respect to the mean
evolution, i.e.,\ $D_m>0$. As the nova shell carries the mean specific
orbital angular momentum
of the WD we also have $D_c < D_m$, hence $\alpha < 2$ (see the Appendix).
The classical stability limit for conservative mass transfer ($D_c=0$)
is at $\alpha = 0$,
while $\alpha < 0$ describes systems where mass transfer would be
dynamically unstable in the absence of nova outbursts.

As can be seen from Fig.~2 the character of
$w(x)$ changes at $\alpha=1$ (note that $w(1) = 1$ in all cases):
\begin{itemize}
\item
If $2 > \alpha > 1$ the function $w(x)$ reaches a maximum at some
$x=x_0$, and $w(x) \rightarrow 1$ for $x \rightarrow \infty$.
We have $w \leq w(x_0) < 2$ in all cases. For the limiting case
$\alpha = 1$, $w(x)$ is monotonically increasing and approaches
the value $w=2$ asymptotically. We conclude that for $\alpha \geq 1$
the mass transfer rate spectrum is always narrow.
\item
If $\alpha < 1$, $w(x)$ is monotonically increasing without bound. The
slope of $w(x)$ increases with decreasing $\alpha$. We have $w(x) = x$
for $\alpha = 0$. The condition $w>10$ can be met in all cases, for any
$x > x_{\rm crit}$, but $x_{\rm crit}$ is a function of
$\alpha$ (shown in Fig.~3). As $\alpha$
is in turn a function of mass ratio and adiabatic stellar index, the
requirement $\alpha < 1$ is equivalent to a condition of the form
$q>q_{\rm crit}$. A good approximation is
\begin{equation}
q_{\rm crit} \simeq 0.45 + 0.4 \left( \zeta + \frac{1}{3}
\right).
\label{qcrit}
\end{equation}
\end{itemize}

\begin{inlinefigure}\label{fig4}
\plotone{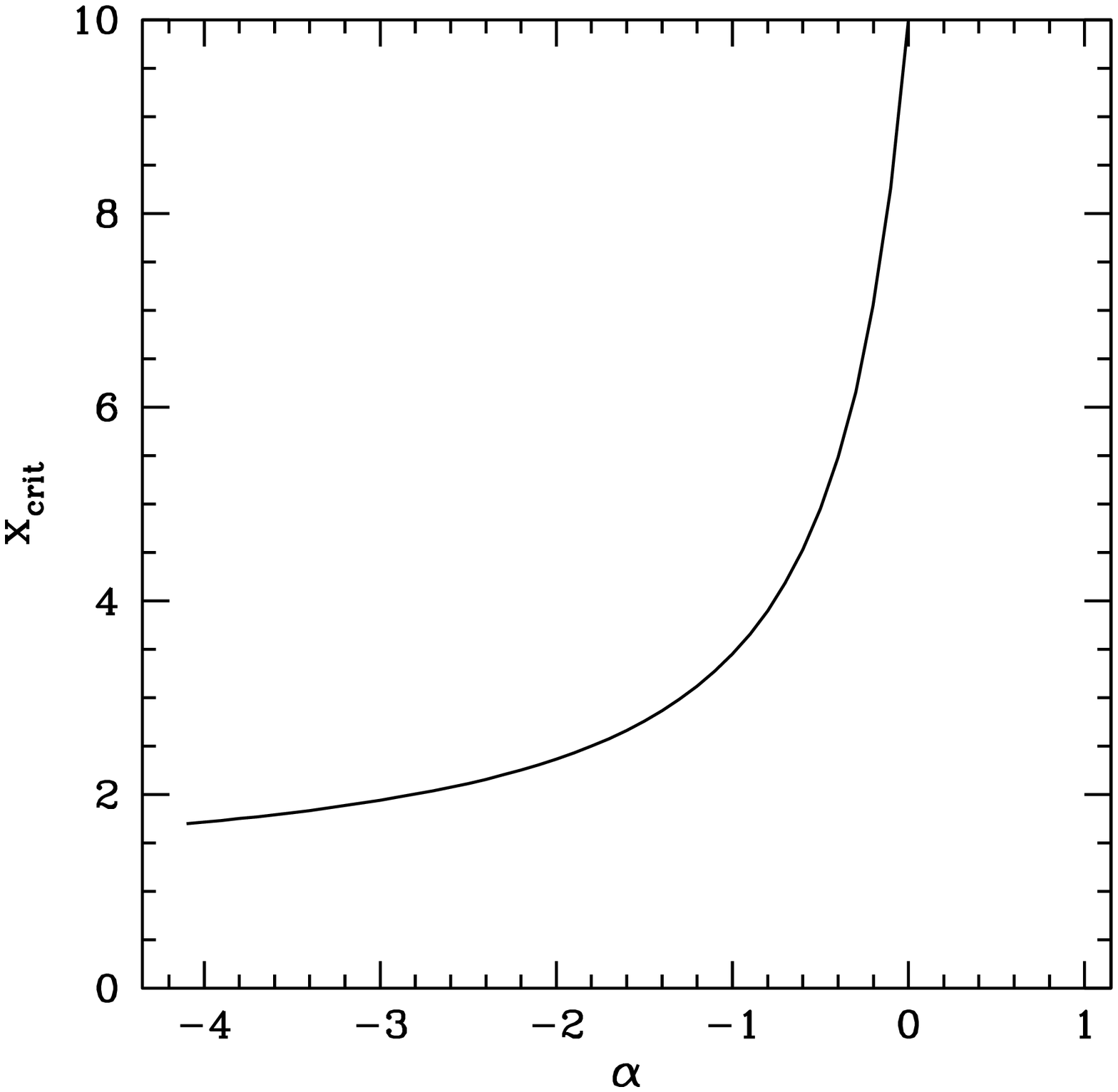}
\caption{The minimal mass accumulation factor
$x_{\rm crit}(\alpha)$ to obtain $w(x)>10$.} 
\end{inlinefigure}

\begin{inlinefigure}\label{fig5}
\plotone{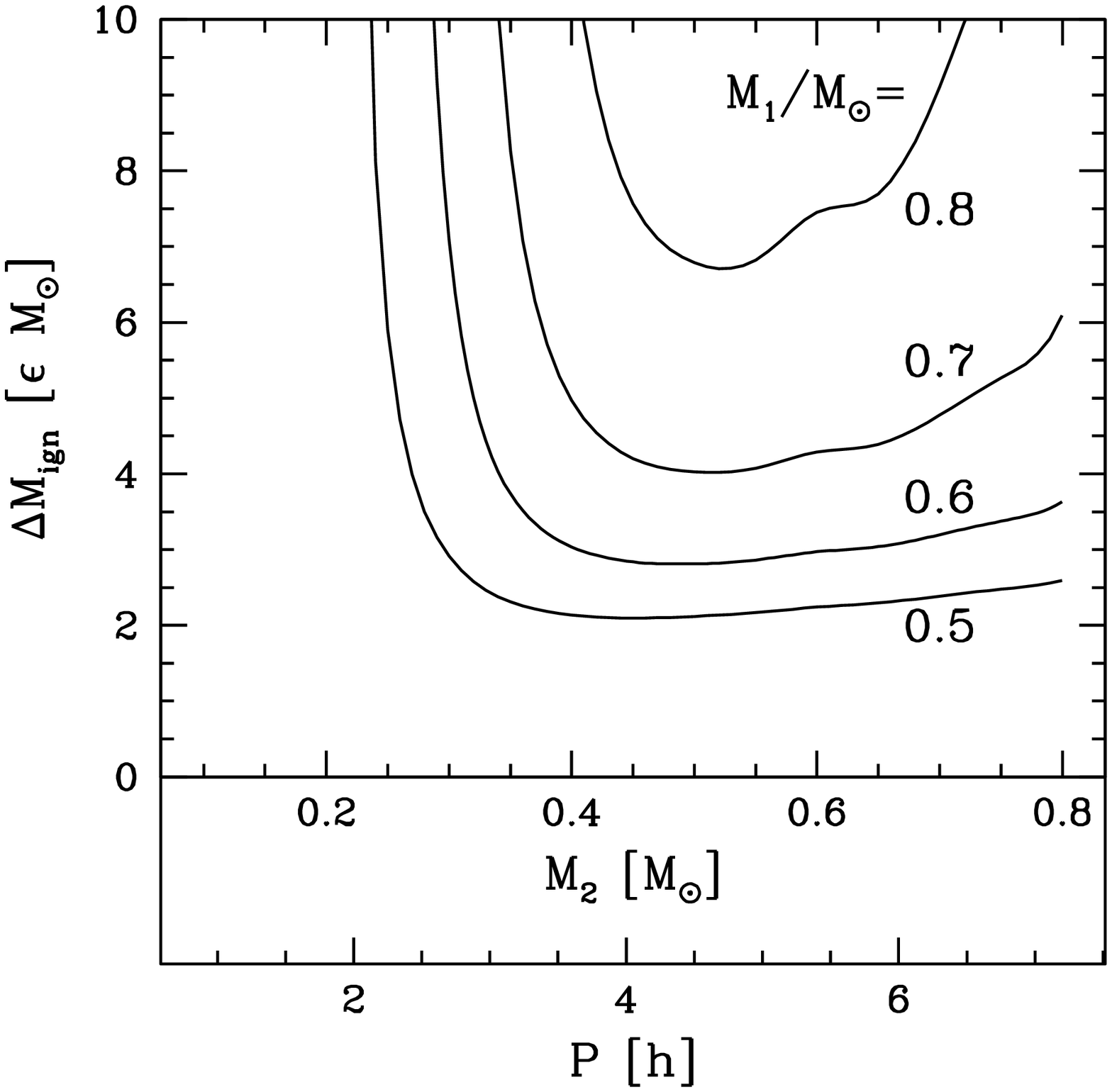}
\caption[dmign1.ps]{Lower limit on the nova trigger mass $\Delta
M_{\rm ign}$ for nova--induced widths of the mass transfer rate
spectrum in excess of a factor 10. $\Delta
M_{\rm ign}$ is in units of $\epsilon\simeq10^{-4} \msun$ and shown as a
function of donor mass $M_2$ or orbital period $P/{\rm hr} = 9
(M_2/\msun)$, for different white dwarf masses $M_1$, as labelled.}
\end{inlinefigure}

The critical conditions $q>q_{\rm crit}$, $x > x_{\rm crit}$
signaling ``wide'' mass transfer rate distributions
becomes more transparent when expressed as constraints on physical
system parameters. For a given WD mass the secondary mass
$M_2$ has to be large enough so that the system is not too far from
mass transfer instability, and the nova outburst trigger mass has to
be larger than a certain lower limit, $\Delta M_{\rm ign} > \Delta
M_{\rm lim}$. Specifically, the former requirement is satisfied when $q >
q_{\rm crit}$, where $q_{\rm crit}$ is given by equation (\ref{qcrit}) and
is typically within a factor of $\sim 1.5$ (smaller than) the value of
$q$ required for mass transfer stability in the absence of
nova explosions. Both requirements are summarized in Fig.~4 where
we plot
$\Delta M_{\rm lim}$ (in units of $\epsilon \msun$) as a function of
$M_2$ for various $M_1$.
In calculating these limits we assumed that $\gamma=1$, i.e.,\ the mass
of the ejected nova envelope is equal to the mass accreted since the
last nova outburst (this affects $D_m$)
and that the secondary's adiabatic mass--radius index $\zeta$ is
the same as for a ZAMS star with mass $M_2$ (shown in
Fig.~5). The figure also gives a rough
estimate of the orbital period the corresponding CV would have,
$P/{\rm hr} \simeq 9 (M_2/\msun)$.

\begin{inlinefigure}\label{fig2}
\plotone{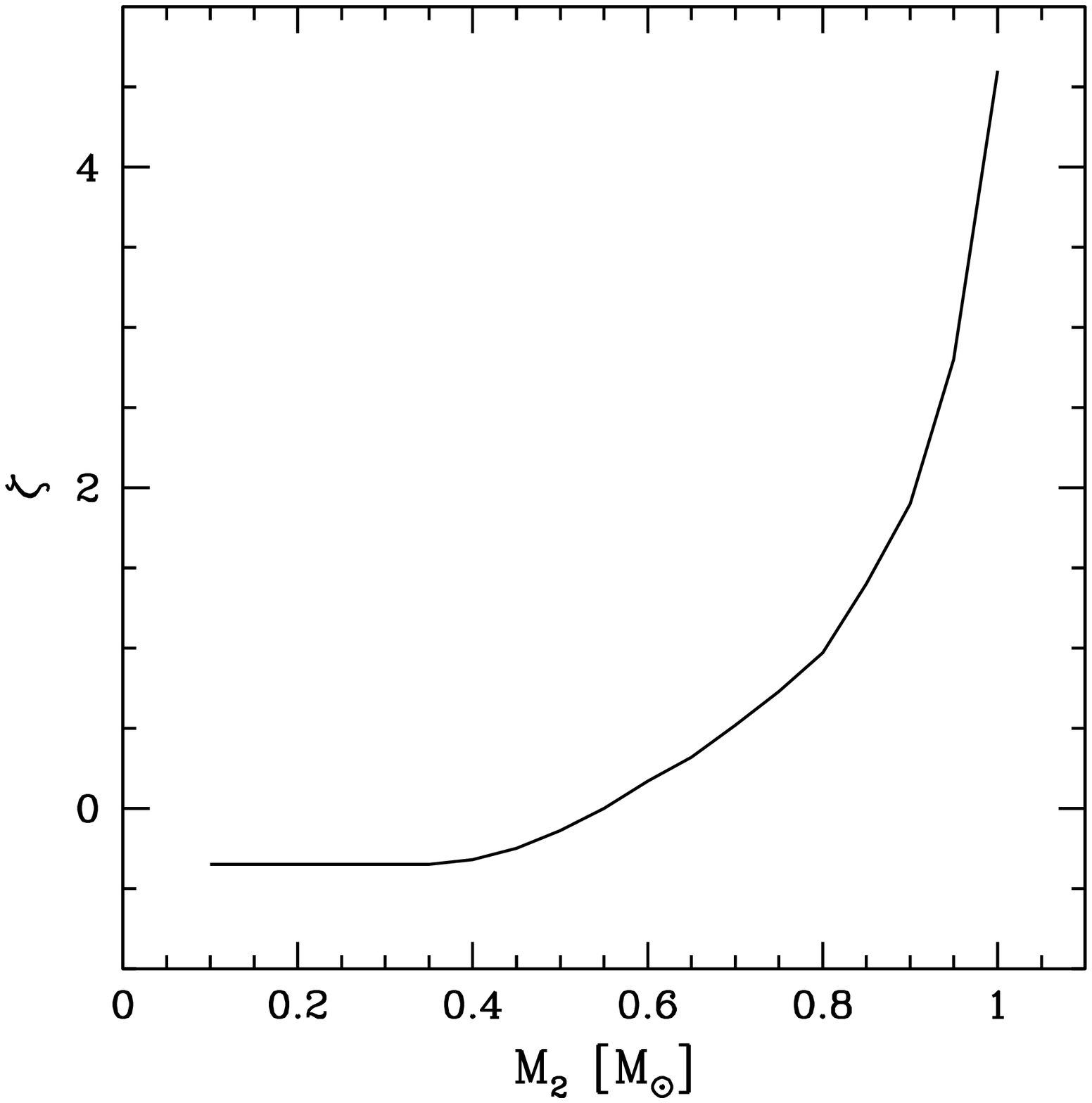}
\caption[zeta.ps]{The adiabatic stellar mass radius index $\zeta$
as a function of stellar mass, for Population I ZAMS stars (from
Hjellming 1989).}
\end{inlinefigure}

The main conclusions to be drawn from Fig.~4 are:

The nova--induced width of mass transfer spectra is {\it in}significant
below the CV period gap ($P\la2$~hr). These short--period systems are
not close enough to instability, i.e.,\ they do not conform to the mass
ratio criterion $q \ga 0.5$. For the same reason the nova--induced
width is negligible whenever the white dwarf is massive ($M_1 \ga 1
\msun$), whatever the orbital period.

Significant nova--induced widths to the mass transfer rate spectrum
can occur above the period gap, for systems with C/O white dwarfs
of mass $0.55-0.8 \msun$ and secondaries
with mass close to the stability limit $M_2 \la 0.7 M_1$. A necessary
requirement is, however, that nova trigger masses are in excess of a
few times $10^{-4} \msun$. The actual value depends on the
photospheric scale height $H$ of the donor star, and {\em increases} with
increasing white dwarf mass.
It is interesting to note that theoretical outburst calculations
unanimously predict that the trigger mass {\em decreases} with increasing
white dwarf mass. Hence we might expect that in a realistic CV population
only those C/O white dwarfs with the smallest mass contribute to an
effective widening.

Chiefly, significant widths in $\dot M$ can
occur above the period gap, but should be least pronounced immediately
above the gap and for long $P$.

In the next section we demonstrate the analytic findings in full
population synthesis models.

\section{Population synthesis models}

In the previous sections, we have shown with an analytic model how
the mass transfer rate in a CV evolves with time during the intervals
between nova explosions (eq. 14).  We then further derived
analytic expressions for the corresponding probability density distribution
in $\dot M$ (eqs. 16 and 17), and the half width of this distribution
(eq. 22).  These expressions are straightforward to evaluate for a given binary
system in a particular evolutionary state.  The behavior of the probability
density distribution is a sensitive function of the binary properties,
including the mass ratio and the stability ratio (eq. 20); this dependence
is illustrated
in Figures $2-5$.  In order to compare these results with observational data
we would like to know what the width in $\dot M$ is as
a function of orbital period for the {\it ensemble} of CV systems that
populate the Galaxy.

In order to explore how broad the range in $\dot M$ would be for an
observed collection of CVs with different evolutionary histories and binary
parameters, we have carried out a population synthesis study with the
effects of nova explosions incorporated.  The basic approach to the
population synthesis is discussed in detail in Howell, Nelson, \& Rappaport
(2001, HNR; see also Kolb 1993, and Kolb 1996, 2001 for reviews).
We start with a large sample ($\sim 3 \times10^6$) of primordial
binaries whose properties are chosen via Monte Carlo
techniques.  The primary mass is chosen from an assumed initial mass
function (IMF; Miller \& Scalo 1979; Eggleton 2001).  We adopt
a distribution of mass ratios for primordial binaries, $p(q)$, which is
approximately flat (see, e.g., Duquennoy \& Mayor 1991).  The initial
orbital period, $P_{\rm orb}$, is chosen from a function which is
constant in log $P_{\rm orb}$ (see, e.g., Duquennoy \& Mayor 1991;
Abt \& Levy 1985).

The evolution of these systems is followed to see which ones
undergo a common envelope phase.  In such events, the envelope
of the evolved giant primary engulfs the secondary, leading to a spiral-in
episode which leaves the secondary in a close orbit with
a white dwarf.  We adopt the
standard approach of assuming that some fraction, $\alpha_{CE}\sim1$,
of the initial orbital binding energy is deposited into the CE as frictional
luminosity (see, e.g., Meyer \& Meyer-Hofmeister 1979;
Sandquist, Taam, \& Burkert 2000).  Primordial
binaries that are too wide will not undergo any significant mass transfer and
will not lead to the formation of CV systems - such wide systems are
discarded in the present study.  Successful pre--cataclysmic variable
systems to emerge from the first
part of the population synthesis calculations are those which do undergo a
common envelope phase and yield a close binary consisting of a white dwarf
and low mass ($\lesssim~1~M_\odot$) companion.

The second part of the population synthesis considers those white-dwarf
main-sequence binaries for which systemic angular momentum losses,
or a modest amount of evolution by the normal companion star, can
initiate Roche-lobe contact within a Hubble time.  At the start of mass
transfer from the original secondary to the white
dwarf the mass transfer may be driven by systemic angular momentum
losses, or if the secondary has a mass which is too large, the transfer may
proceed on either a thermal or dynamical timescale.  We do not consider
either of these latter two cases, although the former most likely leads to
a ``cousin"
of CVs, the supersoft X-ray sources (see, e.g., Kahabka \& van den Heuvel
1997).  Tests for stability at the onset of mass transfer and all other phases
of the CV evolution are done in the binary evolution code by requiring that
$D$ as defined in equation (8) is always positive.  The binary evolution code
is the same as that used by HNR which, in turn, is a descendant of the code
developed by RVJ, and is based on a bipolytropic stellar structure.  Each of
the CV systems is evolved in detail through the mass-transfer phase
(CV phase) until the donor star has been reduced to a negligible mass
(typically 0.03 $M_\odot$).   The expression for magnetic braking, which
is the principal driver of mass transfer for systems above the period gap, is
taken from RVJ and Verbunt \& Zwaan (1981) and adopts values for
RVJ's parameters of $\gamma_{\rm RVJ} = 3$ and $f = 1.0$.   In a
typical population synthesis run, some $\sim 2 \times10^4$ such
binaries are evolved through the CV phase.

We have modified the treatment of mass transfer in the bipolytrope
code so that we can explicitly follow the movement of the Roche lobe
through the atmosphere of the donor star as the system evolves between
nova explosions.  A few illustrative binary evolution runs were made with this
modified code (these are not used or presented in this paper).   However,
this type of explicit calculation requires very short timesteps (sometimes
as short as years), and is therefore very computationally intensive for use
in a population synthesis study.  Fortunately, the binary system and donor star
properties do not evolve significantly during a single interval between nova
explosions, and therefore much of the analytic formalism developed in this
paper can be simply carried over to the population synthesis calculations.  In
fact Schenker et al. (1998) showed explicitly that CV evolution with the
appropriate steady mass-loss rate approximates, to a high degree, the
average evolution of a CV with discrete nova mass loss events.  Thus, in
the population synthesis study we utilize a steady mass loss rate wherein
the ejected matter in nova explosions carries away the specific angular
momentum of the white dwarf.

For each time step in the evolution of a particular CV we store the amount
of time the system spends within a given range of $\dot M$ and $P_{\rm orb}$.
However, for each time step, in addition to recording the actual computed
value of $\dot M$ which represents the case of continuous mass loss, we
utilize equations (14) and (17) to tell us how $\dot M$ is distributed in the
nova-evolution case. We take from the binary evolution code instantaneous
parameters such as the stellar masses, atmospheric scale height, average time
to accrete an ignition mass, adiabatic stellar index, and so forth, i.e.,
all the parameters necessary to evaluate equations (14) and (17).

\begin{figure*}
\centerline{\epsfxsize=1.0\hsize{\epsfbox{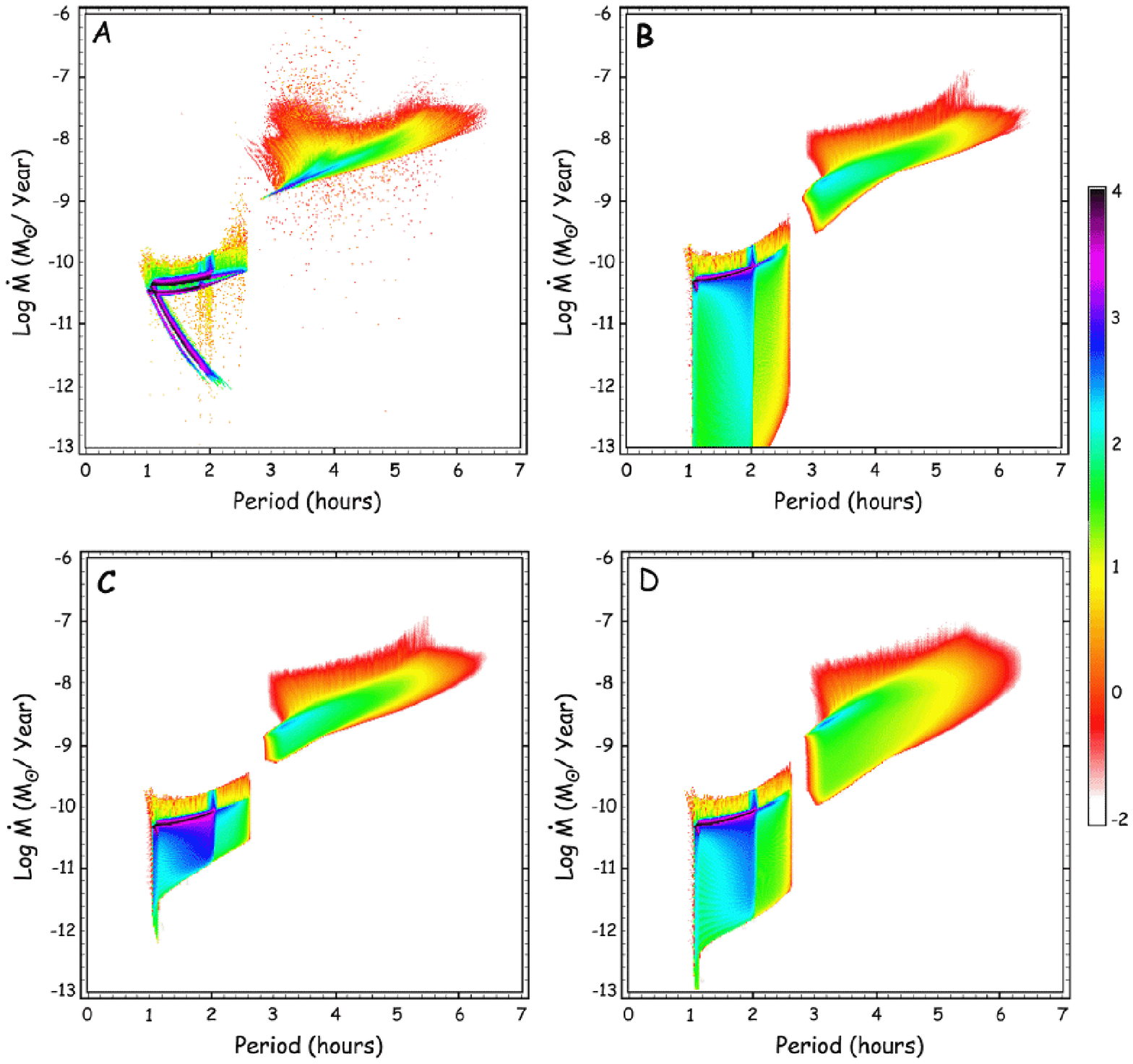}}}
\caption[color.ps]{Probability density distribution (PDF) of a CV
population in the orbital period - mass transfer rate plane. The
color-code refers to log(PDF). See text for details.
{\em Panel A}: Mean evolution, no novae.
{\em Panel B}: Effect of novae taken into account, trigger masses given
by eq.~\ref{dmtrigger}.
{\em Panel C}: Effect of novae taken into account, trigger mass $1
\times 10^{-4} \msun =$~const.
{\em Panel D}: Effect of novae taken into account, trigger mass $3
\times 10^{-4} \msun =$~const.}
\label{fig6}
\end{figure*}

\begin{figure*}
\centerline{\epsfxsize=1.0\hsize{\epsfbox{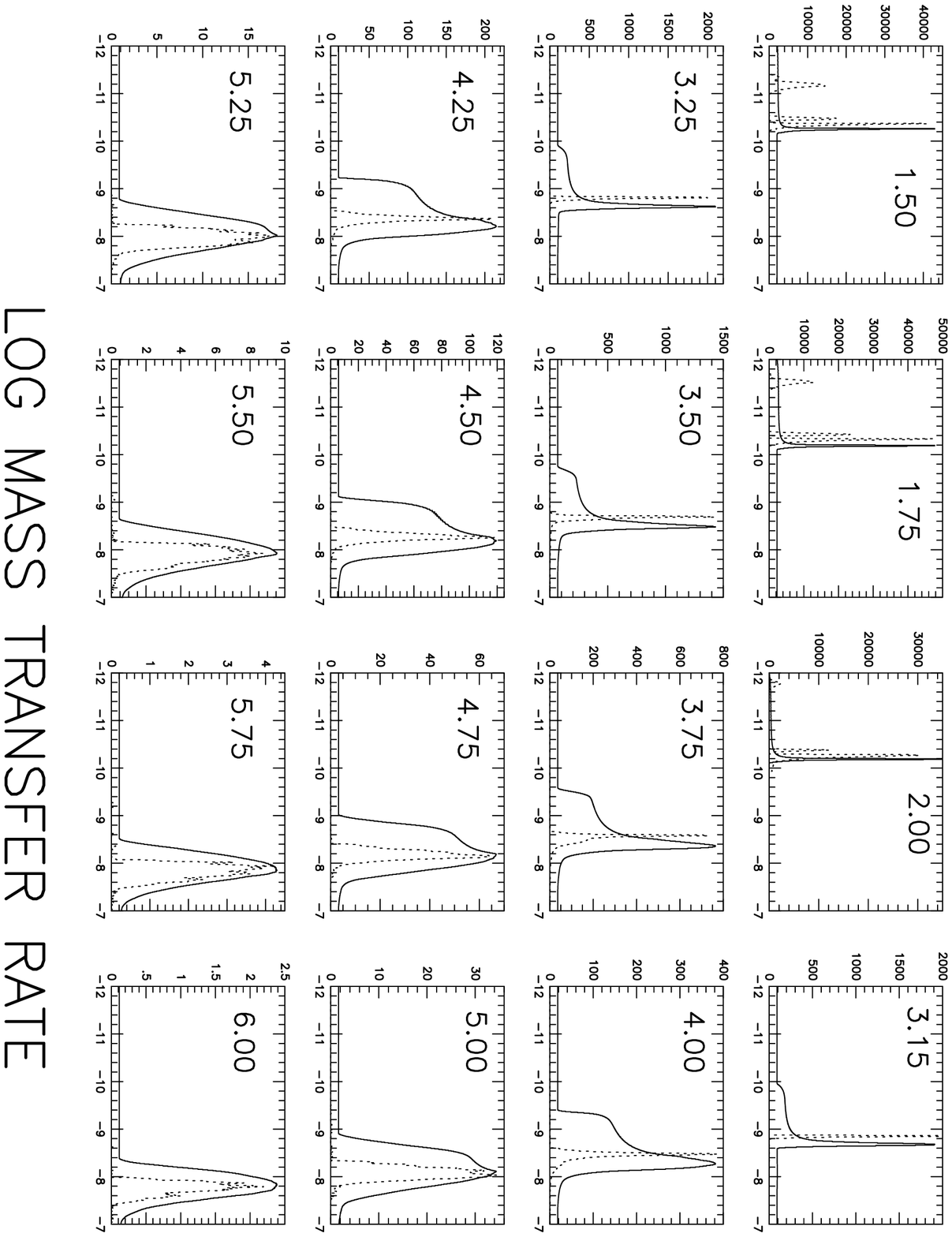}}}
\caption[hist.ps]{Slices in orbital period of the PDFs shown in
Fig.~6, on a linear scale. Solid curves: trigger mass $3 \times
10^{-4} \msun$ (Fig.~6d). Dashed: no nova case (Fig.~6a). Each panel
is labelled with the corresponding orbital period in hrs.}
\label{fig7}
\end{figure*}

The results are shown first in the $\dot M-P_{\rm orb}$ plane, color coded
according to the logarithm of the probability of finding a system in a
particular part of this parameter space.  In Figure~6a the results
are shown for the case of a steady mass loss instead of the more appropriate
discrete nova mass-loss events (see also HNR).  Note that for systems above
the period gap the half width of the distribution in $\dot M$ at any given
value of $P_{\rm orb}$ is no greater than a factor of $\sim 2$.  In
Fig.~6b the same results are shown for the case of discrete nova
events, where the ignition mass is taken to be
\begin{eqnarray}
\Delta M_{\rm ign} & \simeq & 4.4 \times 10^{-4} \, \msun \times
\nonumber \\
& & {\left( \frac{R_1}{10^9 {\rm cm}} \right)}^4 \,
{\left( \frac{M_1}{\msun} \right)} \,
{\left( \frac{\dot M}{10^{-9} \msunyr} \right)}^{-1/3}
\label{dmtrigger}
\end{eqnarray}
($R_1$ is the white dwarf radius). This is an approximate
analytic fit (cf.\ Kolb 1995) to results from
fully time-dependent thermonuclear runaway (TNR) calculations for cold
(old) white dwarfs by Prialnik \& Kovetz (1995), which we use for
illustrative purposes only.
For the WD radius we used the approximation by Nauenberg (1972)
\begin{equation}
\frac{R_1}{R_\odot} = 0.0112 {\left\{
{\left( \frac{M_1}{1.44 \msun}  \right)}^{-2/3}
- {\left( \frac{M_1}{1.44 \msun}  \right)}^{2/3}
\right\}}^{1/2}.
\end{equation}

We also tested other expressions that fit TNR calculations reported
in Nomoto 1982, and semi-analytic solutions to the thermal equilibrium
behavior of H-rich shells given by Fujimoto (1982). The population
synthesis results are quite similar in all cases. In the results leading
to Fig.~6b all CVs with He white dwarf accretors have been
eliminated.  This is due to the uncertainties in our understanding
of the stability and properties of the nuclear burning
on the surfaces of such low-mass white dwarfs.  Note that for the
remaining systems, most of which would have CO white dwarfs, the width
of the $\dot M$ distribution at a given $P_{\rm orb}$ is somewhat broader
than in the case of CV evolution without novae (e.g., Fig.~6a).  For
systems above the period gap the full width at 50 percent probability of
the $\dot M$ distribution is about a factor of 3, but is still not nearly
broad
enough to explain the observational data on CVs.  By contrast, for systems
below the period gap, the width looks quite
broad due to the logarithmic color coding;
however, a more quantitative look (see also
Fig.~7) will show that most systems would have values of
$\dot M$ closely bunched around the values found in the no-nova case.
The long colored tail which appears toward lower $\dot M$ has quite
small probability densities (see e.g.\ Figure~1),
consistent with the findings of Section~3.

Finally, in Fig.~6c and Fig.~6d we show the
results for the somewhat {\it ad hoc} assumption that all novae burn
and eject $1 \times 10^{-4} M_\odot$ or $3 \times 10^{-4} M_\odot$ of
matter.   Fig.~6c, for  $dM_{\rm ign} = 1 \times 10^{-4}
M_\odot$, shows a rather uniform width in $\dot M$ (above the period gap)
of about a factor of 3, which is comparable with that for the case where
$dM_{\rm ign}$ is calculated from equation (\ref{dmtrigger}) (see
Fig.~6b).  In contrast, the distributions in Fig.~6d,
which were computed for the case where $dM_{\rm ign} = 3 \times 10^{-4}
M_\odot$,
show a width in $\dot M$ that is nearly a factor of
10, in reasonable agreement with the observations.

The color-coded images provide a qualitative overview of the evolutionary
properties expected for CVs in the $\dot M-P_{\rm orb}$ plane.  However,
for a more quantitative view we take slices in orbital period and plot
ordinary linear histograms of the distribution in $\dot M$ for each of 16
intervals of $P_{\rm orb}$.  The results corresponding to the color image of
Fig.~6d are shown in Fig.~7 as solid curves.  The dashed
curves (narrower distributions) are for the no-nova cases and are shown for
purpose of comparison.  In each panel, the interval in $P_{\rm orb}$ is
typically 1/4 hr wide, and centered on the value indicated in that panel.
A perusal of this figure confirms the description of the width in $\dot M$
for various regions of $P_{\rm orb}$ as discussed in connection with the
color figures in the $P_{\rm orb}-\dot M$ plane, and the analytic
expressions for the mass transfer rate spectrum in section 3.3.

We note that for donors with mass $\ga 0.6~ \msun$ the bipolytrope models
underestimate the adiabatic stellar index. This might cause a slight
overestimate of the width of the mass transfer rate spectrum at periods
longer than about 5 hrs.

\section{Discussion}

We have investigated the width of the distribution of mass transfer rates
for CVs with similar orbital period, induced by nova outbursts, as a
function of orbital period. We used a semi--analytical model that
describes the time evolution of the mass transfer rate between
outbursts to probe the parameter space.
A detailed population synthesis model demonstrates the
overall effect on a population of CVs with standard assumptions about
formation and evolution.

CVs where mass transfer is close to instability or unstable, in the
absence of nova outbursts, but stabilized through outbursts,
contribute most to an effective widening of the $\dot M$
distribution.
Hence the effect is most pronounced where this instability
occurs, i.e.,\ in systems with large mass ratio $M_2/M_1$.
For this reason the effect is insignificant below the period gap
(despite possibly large ratios of maximum to minimum values of
$\dot M$ after and prior to nova explosions), and
at long orbital periods where all CVs are dynamically stable.
In the intermediate period regime, chiefly between 3 and 6 hrs,
a surprisingly large fraction of CVs are close to the optimum
parameter range that maximizes the impact of novae on the spread of
$\dot M$. The main contribution is from systems with white dwarf
masses less than $0.8 \msun$.

We find that the spread of mass transfer rates is significant,
i.e.,\ covering an order of magnitude, only if the mass accreted
between two outbursts exceeds a few times $10^{-4} \, \msun$. This
critical value increases with increasing white dwarf mass, and is
inversely proportional to the relative
scale height $H/R$ of the donor star's photosphere.

Our goal was to determine the maximum effect of nova outbursts on the
spread of $\dot M$ values. Therefore we assumed in our analysis that
there are no frictional orbital angular momentum losses due to the
orbital motion of the secondary inside the expanding nova envelope.
In that case the mass transfer rate always decreases as a result of an
outburst. If there is dynamical friction, the decrease of $\dot M$ becomes
smaller with increasing friction, and eventually turns into an
increase (see, e.g.,\ Schenker et al.\ 1998). A similarly large amplitude
change in $\dot M$ as in the absence of friction, but in the opposite
direction, requires an implausibly strong dynamical friction.

We assumed further that the mass of the ejected nova envelope is equal
to the mass the white dwarf accretes between two outbursts (i.e.,\
$\gamma = 1$). The effect of changing $\gamma$ is mainly felt through
the stability of the systems. A larger $\gamma$ decreases $\zeta_m$,
see equation~(\ref{zetam}), and hence widens the parameter space available to
systems that are dynamically unstable in the absence of nova
outbursts but stable with outbursts. Hence any $\gamma > 1$ increases
the effect of novae on the spread of $\dot M$, while any $\gamma < 1$
decreases the effect, compared to the case we have investigated in detail.
If $\gamma = 2$ (i.e.,\ the white dwarf ejects twice as much
mass as it has accreted) the limiting accreted mass required to
cause a half width of an order of
magnitude for CVs with a $0.6~\msun$ WD decreases from $3.0 \times
10^{-4} \, \msun$ to $1.5 \times 10^{-4} \, \msun$. At the same time
the curves in Fig.~4 extend to somewhat smaller values of
$M_2$. On the other hand, if $\gamma = 0.8$ (i.e.,\ the white dwarf
retains 20\% of the mass accreted previously) the limiting mass
becomes $3.5 \times 10^{-4} \, \msun$. The curves in Fig.~4
move up, those for higher $M_1$ more so than those for lower $M_1$.
Systems with $M_1 = 0.5\msun$ and $M_2 \ga 0.5 \msun$ are
dynamically unstable even in the presence of nova outbursts.

Another effect that could reduce the impact of nova outbursts on the
mass transfer rate spectrum is wind losses from the disk prior to outburst,
so that the white dwarf accretes less mass than is actually
transferred via Roche lobe overflow. If this mass carries away the
specific orbital angular momentum of the white dwarf the effect is
that the Roche lobe index describing the evolution between two
outbursts is no longer the conservative one, but a smaller one, given
by equation (\ref{zetam}), with $\gamma$ now referring to the fraction
of transferred mass that is lost in the wind. The net effect is to
stabilize the systems, and hence to reduce the parameter space
available to nova--stabilized systems.

Finally, any weighting in favor of bright systems will make the
maximum of the mass transfer rate spectra seen in Fig.~7 more
pronounced, i.e.,\ the distribution of $\dot M$ will be more skewed.
The distributions will therefore appear narrower.

Our results are largely independent of the assumed functional form of
the magnetic braking law. The differential effect of nova outbursts on
the width of the mass transfer rate spectrum is the same for any orbital
braking law. This is significant for the relative distribution of
dwarf novae and novalikes as a function of orbital
period. If the critical rate for stable disk accretion separates these
two classes, then a suitably chosen orbital braking strength, combined
with the nova--induced $\dot M$ spread, can reproduce the observed
relative distribution.

A particularly striking detail of the population synthesis simulations
is the effect on the mass transfer rate
spectrum at periods immediately above the gap (Fig.~7). The
unwidened distribution is narrow and centered on $\dot M \simeq 1.5
\times 10^{-9} \msunyr$. In the standard model, the width of the
period gap essentially fixes this value, and provides a calibration
for the magnetic braking strength (see, e.g.\ Kolb 1996). As noted already
by Shafter (1992), this rate is lower than the critical rate for disk
stability, hence suggesting that just above the gap most CVs should be
dwarf novae. The exact opposite is observed (e.g.\ Ritter \& Kolb
1998). Significantly, the nova--widened $\dot M$ distribution has a
pronounced maximum above the secular mean, at $\dot M \simeq 2.5
\times 10^{-9} \msunyr$ (see panel for $P=3.15$~hrs in Fig.~7),
thus allowing a fraction of systems to
appear as novalikes. Note, however, that the nova widening does not
affect the mean value of the $\dot M$ distribution at any given orbital
period. Hence in the nova--widened distributions shown in Fig.~7
there would still be more dwarf novae than novalikes immediately above
the period gap. This is different if there is an observational
selection effect in favor of brighter systems. If the selection is
strong enough dwarf novae that are present in the intrinsic
population could be effectively suppressed in the observed sample.

A comment on ``hibernation'' is in order. In the ballistic ejection cases
considered here the post nova transfer rate can indeed be much smaller
--- by one or two orders of magnitude ---  than the secular mean rate
(see, e.g.,\ Fig.~1). Generally, after an outburst the systems
evolve back close to the secular mean within a time $f_n t_{\rm rec}$ which
is a fraction $f_n<1$ of the nova recurrence time $t_{\rm rec}$.
At periods where the nova--induced spread of transfer rates is
negligible, e.g.,\ below the CV period gap, $f_n \la 0.1$, while for
systems close to dynamical instability $f_n \simeq 0.5$ (see, e.g.,
Fig~1).
In other words, the majority of CVs spend a significant fraction of
their time close to the secular mean. This is the converse of
the original hibernation model hypothesis $f_n \sim 1$.
As recurrence times are generally very long, $10^4$ yrs
or longer, this is consistent with observations of apparently very
faint post nova systems for the few observed historical novae which
obviously had at most a few centuries to recover from the outburst.
A similar conclusion was reached by Iben, Fujimoto \& MacDonald
(1995; their Appendix B).

Our findings rely on the assumption that the majority of CV secondaries
are unevolved, i.e.,\ main--sequence stars (not necessarily in
thermal equilibrium) where core hydrogen depletion is not yet very
advanced. This is consistent with the standard model of CV formation,
as employed in our population synthesis calculations. There is
evidence, however, that a significant fraction
of CVs do have a somewhat evolved donor star (Baraffe \& Kolb 2000;
see also Kolb 2001). This should be followed up by future studies.

We note that Spruit \& Taam (2001) have discussed the fluctuation of
$\dot M$ in the context of a hypothetical circumbinary disk,
but had difficulties finding a convincing model that would at the same
time preserve the orbital period gap of CVs.

Irradiation (Ritter, Zhang \& Kolb 2000, and references
therein) has been considered as an effect that superposes mass
transfer cycles, on the thermal time of the donor star, onto the secular
mean without changing the mean. Without additional driving mechanisms,
such as the coupling to a consequential angular momentum loss
(e.g.\ McCormick \& Frank 1998) all indications suggest that
irradiation--driven cycles disappear for systems with donor mass
$\la 0.7 \msun$.

We conclude that nova--induced variations of the mass transfer rate
may account for the observed spread of $\dot M$ above the period gap,
but only if the trigger masses on intermediate--mass C/O white dwarfs are in
excess of $\sim 2 \times10^{-4} \msun$. CVs with the highest transfer rates
should be close to mass transfer instability, i.e.,\ should have a fairly large
mass ratio. This constitutes an observational test of the
nova--induced spread of CV transfer rates.

\bigskip

\acknowledgments

We are grateful to Lorne Nelson, Andrew King and Hans Ritter
for useful discussions.
We thank Mike Shara and Goce Zojcheski for their participation
in an earlier phase of this research.
Comments by the anonymous referee helped to improve the presentation
of the paper.
Bart Willems spotted an inconsistent definition of $\beta_2$.
This research was supported in part by PPARC (UK), and by NASA under
ATP grants
GSFC-070 and NAG5-8500 (to SBH), and NAG5-7479 and NAG5-4057 (to SAR).
UK thanks the Aspen Center for Physics (where this work has been
completed) for hospitality and support.

\appendix

\section{Analytical model: further details}

This appendix provides details and derivations for the model developed
in Section 3.

\subsection{The Roche--lobe index}

The stationary mass transfer rate (\ref{Xstat}) depends on the Roche
lobe index $\zeta_L$. We obtain $\zeta_L$ by comparing the definition
(\ref{Rldot}) with the time derivative of
\begin{equation}
J = M_1 \, M_2 \sqrt{\frac{G\,a}{M_b}}
\label{j}
\end{equation}
($a$ is the orbital separation, $M_b=M_1+M_2$ the total binary
mass), and by noting that $\rl/a$ is a function of mass ratio
$q=M_2/M_1$ only. The time derivative of the left hand side of
(\ref{j}) is the sum $\dot J_{\rm sys} + \dot J_{\rm ej}$ of systemic
losses and losses due to mass that is ejected from the system. We
parametrize the latter by
\begin{equation}
\dot J_{\rm ej} = \nu \, \frac{J}{M_b} \, \dot M_b ,
\end{equation}
so that $\nu$ measures the specific angular momentum of the mass lost from
the binary, in units of the mean specific angular momentum of the
orbit. We further assume that the mass lost from the system is a
fraction $\gamma>0$ of the transferred mass,
\begin{equation}
\dot M_b = \gamma \dot M_2 \;\;\;\mbox{or}\;\;\;\dot M_1 = (\gamma - 1)
\dot M_2.
\label{eq:gamma2}
\end{equation}
This is consistent with the definition of $\gamma$ in (\ref{eq:gamma})
if (\ref{eq:gamma2}) is understood as the net effect over one nova
cycle.

Using the notation
\begin{equation}
\beta_2  = \frac{q}{\rl/a} \frac{\dd \rl/a}{\dd q}
\end{equation}
this finally gives
\begin{equation}
\zeta_L \, = \, \gamma \left\{ (2\nu+1)\frac{M_2}{M_b} +
(-\beta_2-2)\frac{M_2}{M_1} \right\} + \zeta_c,
\label{zetal1}
\end{equation}
with $\zeta_c$, the Roche lobe index for conservative mass transfer
($\gamma=0$), given by
\begin{equation}
\zeta_c \, = \, 2 \frac{M_2 - M_1}{M_1} + \frac{M_b}{M_1} \beta_2.
\end{equation}

If $q\la 1$ we have $\rl/a \simeq 0.46(M_2/M)^{1/3}$ (Paczy\'nski 1971),
hence
\begin{equation}
\beta_2 \simeq \frac{1}{3} \, \frac{M_1}{M_b}
\label{beta2}
\end{equation}
so that
\begin{equation}
\zeta_L \simeq \gamma \left\{ 2\nu\frac{M_2}{M_b} + \frac{2}{3}
\frac{M_2}{M_b} - 2 \frac{M_2}{M_1} \right\} + \zeta_c,
\label{zetal2}
\end{equation}
and $\zeta_c$ simplifies to the standard expression
\begin{equation}
\zeta_c \simeq 2 \frac{M_2}{M_1} - \frac{5}{3}.
\end{equation}

\subsection{Mean evolution}

If the nova ejecta carry the specific orbital angular momentum of the
WD we have $\nu = M_2/M_1$, and
(\ref{zetal2}) simplifies to
\begin{equation}
\zeta_m \simeq - \gamma \frac{4}{3} \frac{M_2}{M_b} + \zeta_c.
\label{zetam}
\end{equation}
The corresponding long--term mean mass transfer rate introduced by
(\ref{trec}), the ``isotropic wind--average'', is
\begin{equation}
\dot M_m = \frac{M_2/t_{\rm ev}}{D_m}
\label{Xm}
\end{equation}
with $D_m = \zeta - \zeta_m$, cf.\ (\ref{Xstat}). The equivalent
expression with $D_c = \zeta - \zeta_c$,
\begin{equation}
\dot M_c = \frac{M_2/t_{\rm ev}}{D_c}
\label{Xc}
\end{equation}
represents the conservative stationary rate the system would adopt in
the absence of nova outbursts if $D_c>0$. If the conservative
evolution is dynamically unstable, i.e.,\ $D_c<0$, the quantity $\dot
M_c$ as defined in (\ref{Xc}) is still the relevant quantity to use
in the expression for the time evolution of
$\dot M$, equation~(\ref{Xt}).

In the derivation presented here we assume that the thermal relaxation
term $(\partial \ln R/\partial t)_{\rm th}$ which builds up in the
donor star experiencing a discontinuous mass loss history due to nova
outbursts is the same as in the time--averaged, continuous evolution.
In other words, $t_{\rm ev}$ in (\ref{Xm}) is the same as
in (\ref{Xt}). This has been
confirmed by detailed integrations of the secular
evolution using full stellar models.

\newpage

\subsection{Outburst amplitude}

The outburst amplitude ratio $\dot M_f/\dot M_0$ of the mass transfer rate
($\dot M_f$ is $\dot M$ at $t=t_{\rm rec}$) can be calculated from the ejected
envelope mass. From (\ref{mdot}) we have
\begin{equation}
\frac{\dot M_f}{\dot M_0} = \frac{\dot M(\mbox{pre-nova})}{\dot M(\mbox{
post-nova})} = \exp \left\{ \frac{1}{\epsilon} \left( \frac{\Delta
\rl}{\rl} \right)_{\rm out} \right\},
\label{A1}
\end{equation}
where the index ``out'' refers to the nova outburst. Using (\ref{Rldot})
the total change $({\Delta \rl/\rl})_{\rm total}$ of the Roche lobe
radius over one outburst cycle (outburst, followed by conservative
evolution until the next outburst) can be written as
\begin{equation}
\left( \frac{\Delta \rl}{\rl} \right)_{\rm total} =
\left( \frac{\Delta \rl}{\rl} \right)_{\rm out} +
\left( \frac{\Delta \rl}{\rl} \right)_{\rm inter} =
\zeta_m \frac{\Delta M_2}{M_2} + \left( \frac{\Delta J_{\rm sys}}{J}
\right)_{\rm inter},
\end{equation}
while
\begin{equation}
\left( \frac{\Delta \rl}{\rl} \right)_{\rm inter} =
\zeta_c \frac{\Delta M_2}{M_2} + \left( \frac{\Delta J_{\rm sys}}{J}
\right)_{\rm inter}.
\end{equation}
Here ``inter'' refers to the evolution between two outbursts, and $-\Delta
M_2 = \Delta M_{\rm ign} > 0$ is the mass transferred in this phase.
Hence we have
\begin{equation}
\left( \frac{\Delta \rl}{\rl} \right)_{\rm out} \, = \,
\left( \zeta_m - \zeta_c \right) \frac{\Delta M_2}{M_2} \, = \,
\left( D_c - D_m \right) \frac{\Delta M_2}{M_2}.
\label{dRout}
\end{equation}
Using (\ref{dRout}) in (\ref{A1}) we find
\begin{equation}
\frac{\dot M_f}{\dot M_0} = \exp \left\{ \frac{1}{\epsilon} \, \left(
D_m-D_c \right) \,
\frac{\Delta M_{\rm ign}}{M_2} \right\}
\label{A2}
\end{equation}
With
(\ref{zetam}) for $\zeta_m$
this simplifies to equation~(\ref{A3}) given in Section 3.2.

\subsection{The mass transfer rate spectrum}

The integrated form of the mass transfer spectrum (\ref{dtdX}) is
\begin{equation}
t(\dot M) = \epsilon \, t_{\rm ev} \, \ln \left\{ \frac{\dot M(\dot
M_c-\dot M_0)}{\dot M_0
(\dot M_c-\dot M)} \right\}
\label{tX}
\end{equation}
cf.\ Equation (\ref{Xt}).

To calculate either (\ref{tX}) or (\ref{dtdX}) for systems with
specified masses and secular mean rate we need to find $\dot M_0$. This can
be obtained as follows.
From (\ref{Xt}) at $t=t_{\rm rec}$ we have
\begin{equation}
\frac{\dot M_0}{\dot M_c} = \frac{\dot M_0/\dot M_f - \exp\{-t_{\rm
rec}/\epsilon t_{\rm
ev}\}}
{1-\exp\{-t_{\rm rec}/\epsilon t_{\rm ev}\}}.
\label{X0_1}
\end{equation}
Using (\ref{A2}) to replace $\dot M_0/\dot M_f$ and noting that
\begin{equation}
\frac{t_{\rm rec}}{\epsilon t_{\rm ev}} = \frac{\Delta M_{\rm
ign}/\dot M_m}{\epsilon t_{\rm ev}} = \frac{D_m}{\epsilon} \frac{\Delta
M_{\rm ign}}{M_2},
\label{trick}
\end{equation}
(cf.\ (\ref{trec}) and (\ref{Xm})) we finally find
\begin{equation}
\frac{\dot M_0}{\dot M_c} = \frac{\exp \left\{ D_c \, \Delta M_{\rm
ign}/\epsilon
\,M_2 \right\} - 1}{\exp \left\{ D_m \, \Delta M_{\rm ign}/\epsilon
M_2 \right\} - 1}.
\label{X0}
\end{equation}

\end{document}